\documentclass[]{spie}  

\usepackage{amsmath,amsfonts,amssymb}
\usepackage{graphicx}
\usepackage{wrapfig}
\usepackage{pdflscape}
\usepackage{rotating}
\usepackage{booktabs} 

\usepackage[colorlinks=true, allcolors=blue]{hyperref}

\title{ ComboNet: Combined 2D \& 3D Architecture for Aorta Segmentation}

\author[a,b]{Orhan Akal}
\author[b]{Zhigang Peng}
\author[b]{Gerardo Hermosillo Valadez}
\affil[a]{Florida State University Department of Mathematics, Tallahassee,FL, USA}
\affil[b]{Siemens-Healthineers,Malvern, PA, USA}

\authorinfo{Further author information: (Send correspondence to Orhan Akal)\\Orhan Akal: Applied \& Computational Mathematics PhD Candidate at Florida State University and Image Analytics Intern at Siemens-Healthineers E-mail: orhanakal@gmail.com\\  Zhigang Peng: Senior Staff Scientist at Siemens-Healthineers E-mail: zhigang.peng@siemens-healthineers.com\\
Gerardo Hermosillo Valadez: Principle Key Expert, Head of SHS DI SY D II}

\begin{document} 
\maketitle
\begin{abstract}
3D segmentation with deep learning if trained with full resolution is the ideal way of achieving the best accuracy. Unlike in 2D, 3D segmentation generally does not have sparse outliers, prevents leakage to surrounding soft tissues, at the very least it is generally more consistent than 2D segmentation. However, GPU memory is generally the bottleneck for such an application. Thus, most of the 3D segmentation applications handle sub-sampled input instead of full resolution, which comes with the cost of losing precision at the boundary. In order to maintain precision at the boundary and prevent sparse outliers and leakage, we designed ComboNet. ComboNet is designed in an end to end fashion with three sub-network structures. The first two are parallel: 2D UNet with full resolution and 3D UNet with four times sub-sampled input. The last stage is the concatenation of 2D and 3D outputs along with a full-resolution input image which is followed by two convolution layers either with 2D or 3D convolutions. With ComboNet we have achieved $92.1\%$ dice accuracy for aorta segmentation. With Combonet, we have observed up to $2.3\%$ improvement of dice accuracy as opposed to 2D UNet with the full-resolution input image.
\end{abstract}


\section{INTRODUCTION}

\begin{wrapfigure}{r}{0.25\textwidth} 
\vspace{-0.9cm}
    \includegraphics[height=4.5cm]{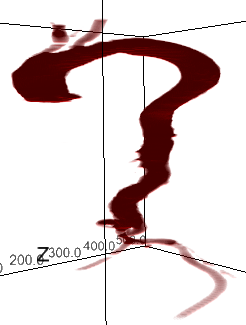}
    \caption{Aorta segmentation with ComboNet}
\end{wrapfigure}

\label{sec:intro}  
Object segmentation has been studied by researchers for many decades now. Among the pioneers of that is the Active contour, Snakes, model\cite{kass1988snakes}, which essentially minimizes and energy function, which is based on the color intensity of the foreground and background. However Snakes model can not handle multiple objects within an image, so in order to overcome this issue, Geodesic Active Contour, GAC, \cite{caselles1995geodesic} model is introduced, which extends Snake model with level sets in conjunction with curve evolution. However, both Snakes and GAC model relies on the existence of a clear edge, which is not always the case. Chan-Vese \cite{chan2001active} tackles this issue by extending Mumford-Shah \cite{mumford1989optimal} energy functional with level sets.

Object segmentation has come a long way with the rise of Deep learning and the help of GPU powered computing, as in other computer vision tasks such as classification, detection, and image-based localization. Especially with the invention of U-Net \cite{ronneberger2015u} and its 3D extension \cite{cciccek20163d}, object segmentation has reached the level of accuracy, which has not been achieved by the aforementioned energy-based models. Primarily because of its encoding and decoding structure and, more importantly, incorporation of ResNet's Residual connection \cite{he2016deep} into their model, not to mention the power of dep learning. Well deserved fame of UNet attracted the researcher to outdo UNet by configuring it such as UNet with Attention Gating \cite{oktay2018attention} and UNet++ \cite{zhou2018unet++}. 

Additional computation complexity that comes with 3D segmentation pushed researchers to segment out sub-volumes of the input then combine the segmentation during postprocessing such as VNet \cite{milletari2016v} and DeepMedic \cite{kamnitsas2017efficient}. DeepMedic combines two 3D segmentation architectures with different resolutions to segment out brain lesion, while segmenting out image patches instead of the entire volume all together to overcome limitations of computation complexity. 

The rise of Deep learning, also urged researchers to revisit energy-based models while combining it with deep learning. For instance, combination of level sets with deep learning \cite{ngo2017combining, hu2018motion} . Also, CVNN\cite{akal2019learning} combined the Chan-Vese algorithm with deep learning in a Recurrent Neural Network fashion.

We tackled the task of segmenting out aorta of a patient's given 3D CT scan. An initial and trivial approach for this problem would have been using a 3D segmentation architecture such as 3D UNet; however, due to GPU memory limitations training such an architecture is not possible unless multiple GPUs are in use. Moreover, passing such an architecture to production is not feasible, given that CT machines have at most one GPU and not necessarily to be state of the art. Thus, 3D segmentation with the full resolution is out of the equation for us. 

A computationally more effective method would have been 3D segmentation with resized 3D input then upsampling the segmentation to the full resolution. However, this method comes with the caveat that once the segmentation is upsampled to the full resolution, the segmentation is crude and not so smooth around the boundary. In order to have more accurate segmentation around the boundary, 2D segmentation with 2D UNet like architecture applied to 2D slices of the 3D volume is a viable option. However, that comes with a caveat too such that 2D segmentation has sparse outliers isolated objects and leakage into a soft tissue, especially where aorta meets heart. In order to achieve accurate segmentation while eliminating gross failure, i.e., sparse outliers and leakage, we combined 2D and 3D segmentation architectures in an end to end architecture called ComboNet. Our method achieves all, besides it beats state of the art and provides production-level speed. Among all, DeepMedic is closest to ComboNet; however, we do not use subpatches, only feed a number of slices at a time. Also we combine 2D architecture with 3D architecture, not two 3D architectures. More importantly, our aim of combining two architecture is not just overcoming computation complexity but eliminating gross failure and finer segmentation at the boundary.

\section{ComboNet Network Design}

ComboNet has three stages; The first two stages are parallel 2D network and 3D network. The third stage is the combiner, where outputs of the first two merge along with the original input image.

\subsection*{2D Network}
2D UNet with $512 \times 512$ input size of axial images, $0.33$mm voxel thickness. The network has 6 Convolution blocks in both encoding and decoding parts. Each convolution blocks have 3 Convolution layers each convolution follows by BatchNorm and ReLU layers. Kernel size (3,3). 

\subsection*{3D Network} 3D UNet with inputs on axial plane sub-sampled four times and on the sagittal plane, no changes made. We used $20$ consecutive slices at a time for 3D segmentation. That yields an input size of $128 \times 128 \times 20$, $1.32$mm voxel thickness on the axial plane. The network has four convolution blocks the same as the 2D network each convolution block has three convolution layers with kernel size (3,3,3), and each convolution layer follows BathcNorm and ReLU layers.

\subsection*{Combiner}
At this stage, outputs of 2D and 3D networks without the sigmoid layer at the end are transferred to the third network. The output of the 3D network up-sampled four times on the axial plane so that it would have the same axial output size i.e., same resolution, as the 2D network. In order for this model work end to end, we fed the same 20 consecutive slices of the 3D network within the same batch of the 2D network. For instance, if the batch size is $4$ for 3D network and there are $20$ consecutive slices in each 3D volume, so for the same iteration batch size for the 2D network will be $4x20=80$. Once the first two stages completed and 3D outputs up-sampled four times on the axial plane there are two options;

\begin{sidewaysfigure}
    \includegraphics[origin=c,height=9.1cm]{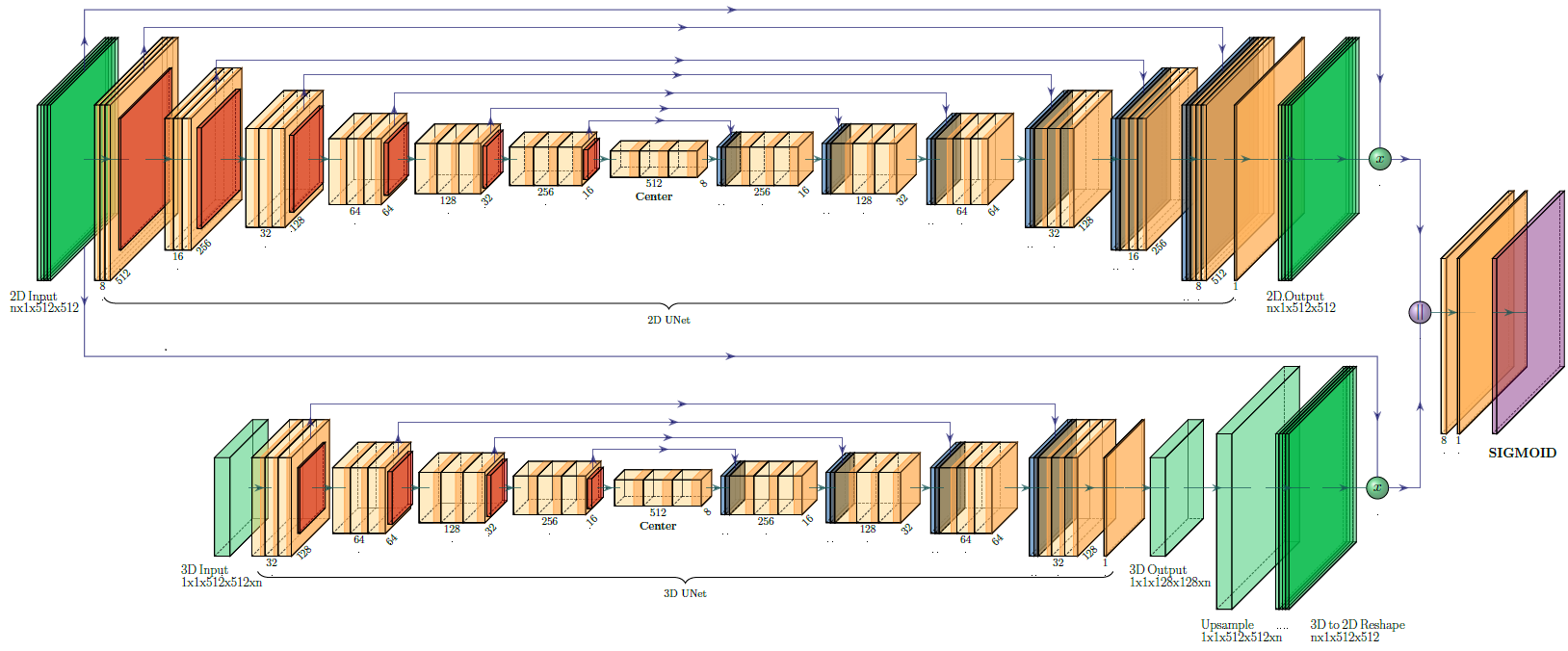}
    \caption{ComboNet-2D architecture. In both ComboNet-2D and ComboNet-3D, 2D UNet and 3D-UNet structures and configuration are the same. The only difference of ComboNet-3D from this architecture is that 2D tiles stacked so that it would have the same shape as up-sampled 3D outputs than concatenated outputs instead of going through 2D convolutions they go through 3D convolutions. }
    \label{fig:architecture}
\end{sidewaysfigure}

\begin{enumerate}
   \item ComboNet-2D: Up-sampled 3D output will be reshaped to have the same size of 2D outputs. i.e., for a batch of $4$ shape of 3D output tensor is $4\times1\times512\times512\times20$ will be reshaped to have a size of $80\times1\times512\times512$.
    \item ComboNet-3D: Similar to the above 2D output will be reshaped to have the same size of 3D output $4\times1\times512\times512\times20$.
\end{enumerate}
In either case, both of the 2D and 3D outputs now have the same size. These outputs will be used as shape priors for the third stage, or one can consider it as heat map activation. These outputs each separately point-wise multiplied with the input image then concatenated together. For ComboNet-2D, this yields an input size of $80\times2\times512\times512$ to the third stage. which then passed through two 2D convolution layers with the BatchNorm and ReLU. Similarly, ComboNet-3D applies two 3D convolution layers.  
The main difference between ComboNet-2D and ComboNet-3D is the convolution type on the combiner part of the network; thus, from this point on, we will refer both of them as ComboNet. ComboNet architecture showed in Figure~\ref{fig:architecture} is from ComboNet-2D. ComboNet structure can work with any 2D and 3D segmentation architectures by replacing 2D U-Net with any 2D segmentation architecture and 3D U-Net with any 3D segmentation architecture.

\section{Optimum Network Architecture Pattern}
\label{sec:optimum}
\begin{table}[]
\centering
\begin{tabular}{lllll}
\toprule
\toprule
Features                              & 2D Network & 2D Network & 3D Network & ComboNet-2D/3D      \\
\midrule
Input size                    & 512        & 256        & 128x20     & (512, 128)x20 \\
Number of Blocks                  & 6          & 5          & 4          & 6,4           \\
Number of Conv. layers per Block            & 3          & 3          & 3          & 3             \\
Feature Scale                 & 8          & 4          & 2          & 8, 2          \\
Central block Size            & 8x8x256        & 8x8x256        & 8x8x256        & 8x8x256,8x8x256           \\
Number of parameters              & 13818297   & 13811569   & 35756321   & 49575473***      \\
Speed (seconds for 250 slices)*           &  0.08      &  0.07225    & 0.09125    & 0.14275       \\
\bottomrule
\bottomrule
\end{tabular} 
\vspace{0.2cm}
\caption{Optimum Network Architectures' features. *Seconds on Titan Xp GPU. It is estimated per slice time x 250 if tested for 2D within a batch of 20 $512\times512$ input and 3D and ComboNet with a batch of 1 with input sizes of $128\times128\times20$ $(512\times512,128\times128)\times20$ respectively. ***ComboNet 3D has 2290 more parameters than ComboNet-2D as it is using 3D convolution kernels as at the combiner part. }
\label{tab:networks}
\end{table}

Original UNet architecture has 4 Convolution blocks with two convolution layers in each block. For each convolution block, we experimented with 2-4 convolution layers and found three convolution layer is optimum. We observed that while keeping the number of filters the same, i.e., scaling down the number of filters by 'feature scale' (Table-\ref{tab:networks}) and using more Convolution blocks and convolution layers, we get better validation performance. More convolution blocks mean more residual connections, and more convolution layers mean more ReLU activation, thus more non-linearity, thus less overfitting. For 2D UNet  We experimented with Convolution Blocks of 4-7 and found for input size of $512\times512$ (2D512) 6 blocks are optimum, and for input size of $256\times256$ (2D256), five blocks are optimum. Similarly, for 3D UNet with input size $128\times128\times n$ (3D128), we experimented with 4 and 5 convolution blocks and found that four convolution blocks are optimum. 

This optimum network structure for various input sizes carries an important pattern; each convolution block means a max-pooling with stride 2; thus, the input is getting smaller 2 times in each direction. Thus 6 convolution blocks would yield 64 times smaller input to the central block of UNet. As can be seen in Table~\ref{tab:networks}, each of these optimum networks gives an output size of $8\times8\times512$ from the central block of UNet and, more importantly, the smallest residual connection has a size of $16\times16\times256$ anything less than that seems to reduce the accuracy. That is said, the number of filters in the central block is the same for all three networks; 2D512, 2D256, 3D128. Not Only the central block but also all 4 convolution blocks from each side of the central block that is closest to the central block has the same number of filters. For instance, 2D512 and 2D256 are the same except 2D512 has 1 additional convolutional block at the beginning and the end of the network.

\section{Training workflow}
Training a network this complicated is not a trivial task so that we will lay down the training phase. We have trained the network in 4 steps with an optional fifth step; (i) training 2D UNet with full resolution image, (ii) training 3D UNet with low-resolution 3D inputs. (iii) Outputs of 2D UNet and 3D UNet without last sigmoid layer are saved and fed as input to the combiner portion of the ComboNet, i.e., the last two layers. With this step, the training procedure is finalized. (iv) Now that we have trained all three sub-networks separately, we combine them in the ComboNet framework so that the network works end to end and passes to the testing phase; if not, the following optional step is used.

(v) ComboNet that is derived at (iv) can be trained further; however, special care needed. For instance, if the entire network is trained with the same learning rate, the odds are the best performance of (iv) may not necessarily be achieved. Thus, we experimented using a 10-100 smaller learning rate for 2D and 3D UNet portions than the combiner portion. In this way, now it is more feasible to train the network end to end and achieve similar performance to (iv). In order to outperform the performance of (iv) cluster of GPUs needs to be used because if the ComboNet trained end to end largest batch size that we can have is 1-2 on a single GPU, which is very problematic because a network trained with batch size 1-2 would not be able to generalize well. Thus, if a cluster of GPUs used the network can be trained with increased batch size (8 is sufficient), this step would carry the potential to outperform the previous step. We experimented with using smaller learning rates for 2D and 3D subnetworks, which showed potential, yet since we did not experimented with a cluster of GPUs, this step did not outperform the previous step. If one experiments this step with a cluster of GPUs, the accuracy would definitely be increased. That is said the results given on Table-\ref{tab:results} is from (iv) not (v).

Due to severe overfitting, we used Combo loss\cite{taghanaki2019combo} \eqref{eq:comboloss} which is weighted combination of weighted Binary Cross Entropy \cite{ronneberger2015u} and Dice loss \cite{sudre2017generalised}. 
\begin{equation}
L(\beta)=\alpha\left(-\frac{1}{N}\sum_{i=1}^N \beta t_i  \ln p_i-(1-\beta) 
(1-t_i)  \ln(1- p_i)\right)
-(1-\alpha)\frac{\sum_{i=1}^N t_i p_i+s }{\sum_{i=1}^N t_i + \sum_{i=1}^N p_i+s }
\label{eq:comboloss}
\end{equation}
where $t_i$ is ground truth and $p_i$ is ComboNet output for $i^{th}$ pixel, and $0< \alpha, \beta<1$ are paramters to be optimized. For our experimetation we optimized $\alpha=0.4$ and $\beta=0.85$. Also, as learning rate scheduler we used augmented cosine annealing waves \cite{akal2018distributed}.
\section{RESULTS}

At this stage, data is ready weights, and the network is loaded depending on the number of slices n and memory capacity of the GPU, the data can be tested all at once or by dividing into sub-volumes. For instance, we experimented with Titan Xp GPU, which has 12 GB memory at a time it can handle 120 slices when the axial dimension of 2D input is $512 \times 512$. That is said, for this case if there are more than 120 slices than the input needs to be divided on the z-direction into sub-volumes than do the testing and put together. Note: sub-volumes do not need to be overlapping; thus, there is no need for any stitching needed for putting together segmentations of sub-volumes.

We have tested our method with 4-fold cross-validation, and report the results based on average dice coefficient per slice. Our experiments from Table-\ref{tab:results} show that, expectedly, ComboNet 2D/3D always outperforms 2D network performance as it is designed to better the results of the 2D network. On average, ComboNet 2D/3D outperforms 2D512 by $1.3\%$, and it gives improvement up to $2.3\%$, i.e., fold-4. We have achieved the best performance both for each network with fold-2, and ComboNet-2D yields $92.1\%$, which is an improvement of $1\%$ on top of 2D512 performance.

\begin{table}[t]
\centering
\begin{tabular}{llllll}
\toprule
\toprule 
& 2D Net & 2D Net & 3D Net & ComboNet-2D   & ComboNet-3D   \\
           \hline
Input  & 512        & 256        & 128x20     & (512, 128)x20 & (512, 128)x20 \\
\hline
Fold-1     & 85.29      & 86.9       & 87.22      & 86.1          & 85.35         \\
Fold-2     & 91.11      & 88.9       & 90.61      & 92.15         & 91.81         \\
Fold-3     & 87.2       & 86.76      & 88.6       & 89.1          & 89.4          \\
Fold-4     & 86.9       & 87         & 88.6       & 89.17         & 88.8          \\
\hline
mean       & 87.62      & 87.39      & 88.75      & 88.95         & 88.84         \\
\hline
parameters    & 13.8m   & 13.8m   & 35.7m   & 49.5m      & 49.5m   \\
Speed       &  0.08      &  0.07225    & 0.09125    & 0.14275   & 0.14275\\
\bottomrule
\bottomrule
\end{tabular}
\vspace{2mm}
\caption{4-fold cross validation of various architectures. Results are average dice score per slice. Speed is based on inference time of a volume with 250 slices with given axial input size given on the table.}
\label{tab:results}
\end{table} 

Fold-1 is always performing low because, for this fold, testing has severe pathology cases; however, the training set does not have severe pathology cases, i.e., calcification. However, we believe if the network trained with an equal number of healthy and pathology cases, then the network should be able to segment out both of the cases with similar performance; otherwise network memorizes healthy cases and interpret pathology cases as noise.

\begin{figure}
    \centering
\vspace{0.2cm}
    \includegraphics[height=4.cm]{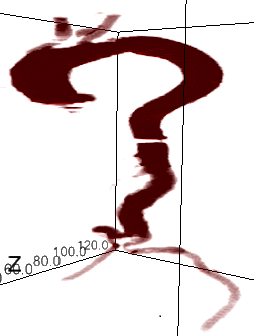}
    \includegraphics[height=4.cm]{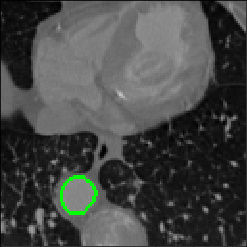}
    \includegraphics[height=4.cm]{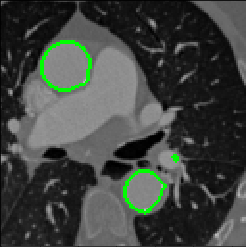}
    \vspace{0.1cm}
    \includegraphics[height=4.cm]{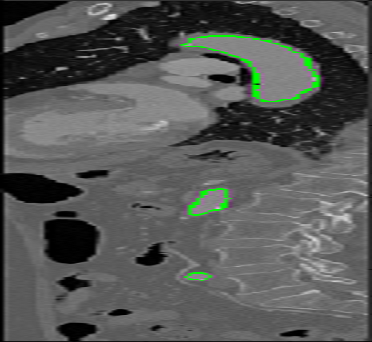}
\vspace{0.1cm}
\hskip 0.001cm
    \includegraphics[height=4.cm]{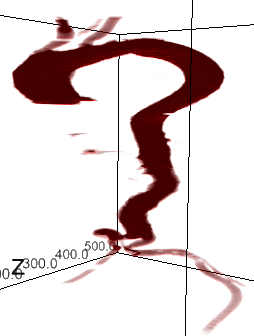}
\hskip -0.001cm
    \includegraphics[height=4.cm]{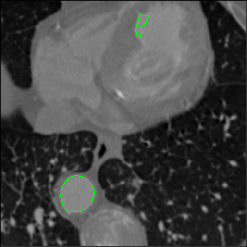}
    \includegraphics[height=4.cm]{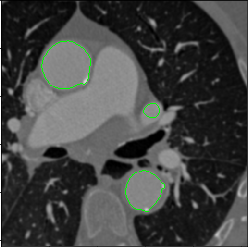}
    \hskip -0.001cm
    \includegraphics[height=4.cm]{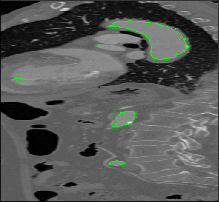}
\vspace{0.1cm}
    \includegraphics[height=4.cm]{pics/segC2D_21_crop.png}
    \includegraphics[height=4.cm]{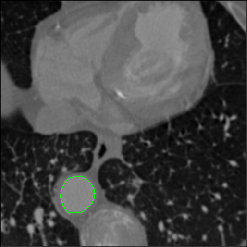}
    \includegraphics[height=4.cm]{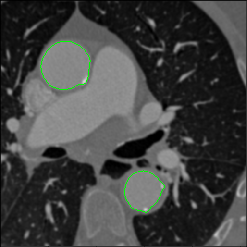}
    \includegraphics[height=4.cm]{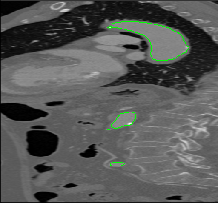}
\vspace{0.1cm}
    \includegraphics[height=4.cm]{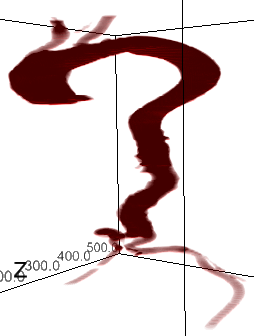}
    \includegraphics[height=4.cm]{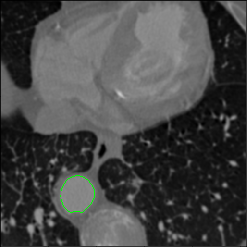}
    \includegraphics[height=4.cm]{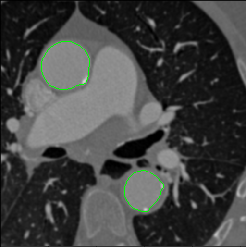}
    \includegraphics[height=4.cm]{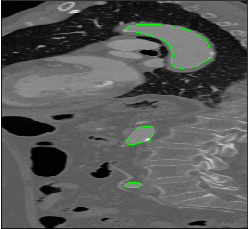}
\vspace{0.1cm}
    \includegraphics[height=4.cm]{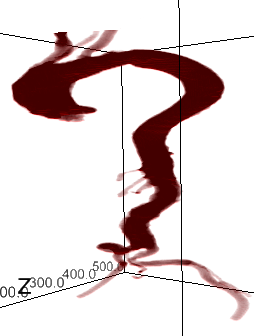}
    \includegraphics[height=4.cm]{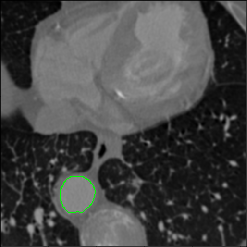}
    \includegraphics[height=4.cm]{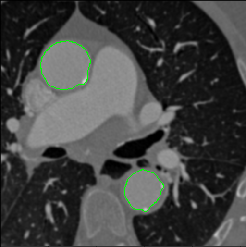}
    \includegraphics[height=4.cm]{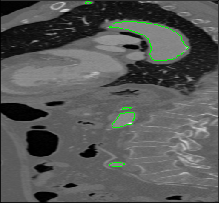}
    \caption{(Top to bottom) Segmentation results of 3D128, 2D512, ComboNet-2D, ComboNet 3D, and ground truth. From left to right 3D view of segmentation, axial view of descending aorta below aorta bridge around heart, axial view at aorta bridge, and sagittal view.}
    \label{fig:snaps}
\end{figure}
Even though we enlisted the results from 2D256 and 3D128, they are not the basis for our performance comparison for ComboNet as they are of smaller input size, thus larger voxel thickness, which gives less fine segmentation. One can see the results from Figure-\ref{fig:snaps} the first row is from 3D128 and second row from 2D512, 3D seems more accurate in terms of having fewer sparse outliers however because of the voxel thickness the boundary of the segmentation cannot be as fine as 2D512. 

Reader can see both from Table-\ref{tab:results} and Figure-\ref{fig:snaps} that ComboNet-2D and ComboNet-3D yields very similar performance and segmentation and always better than 2D512. The first column, first and second row of Figure-\ref{fig:snaps} shows that both 3D128, 2D512 have sparse outliers, and 3D128 has missed the segmentation completely for several layers and partially for other layers and 2D512 also has a tiny bit of hole on the descending aorta. ComboNet 2D/3D eliminates sparse outliers and fixes the holes. Thus, ComboNet 2D/3D promises to yield finer segmentation and to eliminate gross failure.

\subsection{Ablation Study}
\begin{table}[b]
\centering
\begin{tabular}{llllll}
\toprule
\toprule
Experiment ID & Networks & $\#$ Blocks &input   & Loss  function  & Dice  \\
\midrule
$\textit{E}_1$ & ComboNet-2D   &   &  (512,128)x20 & Combo  & 92.15\\
$\textit{E}_2$ & ComboNet-3D   &   &  (512,128)x20 & Combo  & 91.81\\
$\textit{E}_3$ & 2D UNet*      & 6 & 512  & Combo & 91.11 \\
$\textit{E}_4$ & 2D UNet       & 6 & 512  & BCE  $\beta=0.85$ & 90.10  \\
$\textit{E}_5$ & 2D UNet       & 6 & 512  & BCE (no $\beta$) & 87.96 \\
$\textit{E}_6$ & 2D UNet       & 6 & 512  & Dice    & 90.46 \\
$\textit{E}_7$ & 2D UNet       & 6 & 256  & Combo   & 87.87 \\
$\textit{E}_8$ & 2D UNet       & 5 & 512  & Combo   & 90.82\\
$\textit{E}_9$ & 2D UNet       & 5 & 256   & Combo  & 88.90  \\
$\textit{E}_{10}$ & 2D UNet       & 4 & 512 & Combo & 89.05\\
$\textit{E}_{11}$ & 2D UNet-Atten & 6 & 512 & Combo & 88.95 \\
$\textit{E}_{12}$ & 2D UNet-Atten & 5 & 512 & Combo & 87.06 \\
$\textit{E}_{13}$ & 2D UNet-Atten & 4 & 512 & Combo & 87.09 \\
\bottomrule
\bottomrule
\end{tabular}
\vspace{3mm}
\caption{Ablation study results of Fold-2 with various network, input, and loss configurations. (*) These networks are used in ComboNet structure.}
\label{tab:ablation}
\vspace{-9mm}
\end{table}

As noted in Section~\ref{sec:optimum}, we optimized UNet for various input sizes. In this section, we will go over the Dice scores of our choice of network architecture, Network type, and loss function affected the performance. For the sake of simplicity, we will detail mainly around 2D UNet architecture with the results from Table~\ref{tab:ablation}.

First of all, we are comparing the results from UNet and UNet with attention gating for the same configurations. It can easily be seen that UNet is always outperforming UNet with attention gating for the same configuration and same feature scaling, all the while Unet with attention gating has additional attention gating layers and parameters. So that is said UNet with attention gating is less accurate while doing more computations ($\textit{E}_3$ vs $\textit{E}_{11}$, $\textit{E}_8$ vs $\textit{E}_{12}$, $\textit{E}_{10}$ vs $\textit{E}_{13}$). This is contradictory to what the authors of UNet with Attention gating paper promises. We believe their results might be case-specific, or it might be superior when doing multi-object segmentation, which is out of scope of this study.

For input size of $512 \times 512$, optimum UNet architecture has $6$ number of blocks. It can be seen that as the number of blocks decrease from 6 to 4 accuracy decreases ( $\textit{E}_3$ ,$\textit{E}_8$, $\textit{E}_{10}$ ). Similarly for an input size of $256 \times 256$ optimum UNet architecture has 5 blocks instead of 6 unlike the prior(( $\textit{E}_9$ ,$\textit{E}_7$).

Now that we established the grounds for optimum network architecture for input size $512 \times 512 $ is 6 blocks UNet. Let us show the effect of loss function choice. Binary Cross-Entropy Loss alone without any weighting is performing poorly($\textit{E}_5$). When weighting introduced in the BCE i.e., $\beta=0.85$ which penalizes false positives more then the accuracy increases by more than $2\%$ ($\textit{E}_4$ vs. $\textit{E}_5$). On the other hand, if Dice loss is used alone, it performs even better than weighted BCE ($\textit{E}_6$). If we use Combo loss \eqref{eq:comboloss} instead, which is a weighted combination of weighted BCE with Dice loss with weighting parameter choices of $\alpha=0.4$ and $\beta=0.85$, we get the best performance ($\textit{E}_3$). Since we showed that Comboloss with mentioned weights performs better, for the rest of the table, we used the same weighting on the Combo loss. This also confirms what the authors of Combo loss claimed. Needless to repeat, ComBoNet is outperforming all of them.

This ablation study emphasizes the following; (i) choice of the loss function is crucial, especially if there is severe overfitting issue. (ii) An ideal network structure may not necessarily be ideal for all input sizes, especially with UNet, where the output size of the central block is dependent on the input size and the number of convolution blocks. So if the output size of the central block is too small, it will not carry enough information to the encoding path of UNet.

\section{Conclusion}
Product-oriented research requires reasonable accuracy within a reasonable time, i.e., less than 2 seconds per volume, with no gross failure, whereas academic research aims at the highest possible accuracy without considering whether it is suitable for production or not. An ideal architecture that meets academic research might have been 3D UNet with a full resolution image; however, that is not feasible due to computation and memory limitations. Alternatively, if 3D UNet used with the subsampled image, the segmentation is crude when upsampled to the full resolution Figure~\ref{fig:snaps}, so even if it is fast but not accurate enough, does not meet either one of the research purposes. Even though 2D UNet architecture provides high performance, it is not immune to gross failure. On the other hand, ComboNet 2D/3D is the answer to both product and academic research-oriented requirements. It is fast 0.14 seconds per volume(Table-\ref{tab:results}), which is 15 times faster than the upper limit of production and provides a higher accuracy of $92.1\%$, which outperforms state of the art 2D UNet, eliminates gross failure and segmentation precision around the boundary is more accurate and smooth.

\bibliography{references} 
\bibliographystyle{spiebib} 

\end{document}